\documentclass[epsfig,floats,pre,twocolumn,superscriptaddress,
preprintnumbers,floatfix]{revtex4-1}

\usepackage[latin2]{inputenc}
\usepackage{amsmath}
\usepackage{amssymb}
\usepackage{epsfig}
\usepackage{bm}
\usepackage{color}
\usepackage{stackengine}
\usepackage{graphicx}
\usepackage{caption, subcaption}
\usepackage{sidecap}

\begin{document}

\title{Microtubule polymerization generates microtentacles important in circulating tumor cell invasion}
 
\author{Lucina Kainka$^\dagger$}
\affiliation{Department of Experimental Physics, Saarland University, 
66123 Saarbr\"ucken, Germany}
\author{Reza Shaebani$^\dagger$}
\affiliation{Department of Theoretical Physics, Saarland University, 
66123 Saarbr\"ucken, Germany}
\affiliation{Center for Biophysics, Saarland University, 66123 
Saarbr\"ucken, Germany}
\author{Kathi Kaiser}
\affiliation{Department of Experimental Physics, Saarland University, 
66123 Saarbr\"ucken, Germany}  
\author{Jonas Bosche}
\affiliation{Department of Theoretical Physics, Saarland University, 
66123 Saarbr\"ucken, Germany}  
\author{Ludger Santen$^{\ddagger}$}
\affiliation{Department of Theoretical Physics, Saarland University, 
66123 Saarbr\"ucken, Germany}
\affiliation{Center for Biophysics, Saarland University, 66123 
Saarbr\"ucken, Germany}  
\author{Franziska Lautenschl\"ager$^{\ddagger}$}
\affiliation{Department of Experimental Physics, Saarland University, 
66123 Saarbr\"ucken, Germany}  
\affiliation{Center for Biophysics, Saarland University, 66123 
Saarbr\"ucken, Germany} 

\begin{abstract}
Circulating tumor cells (CTCs) have crucial roles in the spread of tumors during metastasis. A decisive step 
is the extravasation of CTCs from the blood stream or lymph system, which depends on the ability of cells to 
attach to vessel walls. Recent work suggests that such adhesion is facilitated by microtubule (MT)-based 
membrane protrusions called microtentacles (McTNs). However, how McTNs facilitate such adhesion and how MTs 
can generate protrusions in CTCs remain unclear. By combining fluorescence recovery after photobleaching 
(FRAP) experiments and simulations we show that polymerization of MTs provides the main driving force for 
McTN formation, whereas the contribution of MTs sliding with respect to each other is minimal. Further, 
the forces exerted on the McTN tip result in curvature, as the MTs are anchored at the other end in the 
MT organizing center. When approaching vessel walls, McTN curvature is additionally influenced by the 
adhesion strength between the McTN and wall. Moreover, increasing McTN length, reducing its bending rigidity, 
or strengthening adhesion enhances the cell-wall contact area and, thus, promotes cell attachment to vessel 
walls. Our results demonstrate a link between the formation and function of McTNs, which may provide new 
insight into metastatic cancer diagnosis and therapy.
\end{abstract}

\maketitle

\section{Introduction}

Circulating tumor cells (CTCs) have attracted widespread attention over the last decade due to their function as seeds for metastasis \cite{AlixPanabieres14,Pantel16,Hamza19,Vermesh18,Ting14,Miyamoto15}. CTCs are transported through the bloodstream \cite{Aceto14,Au16,Sarioglu15}, where they are at high risk of death by apoptosis fragmentation in narrow capillaries, or for recognition by the immune surveillance system. Consequently, less than 1$\%$  of CTCs survive and leave the blood system \cite{Chambers02,Massague16}. The exact mechanism of CTC extravasation is not yet understood, although adhering to blood vessel walls is the first crucial step  \cite{Strilic17}. CTCs can generate long and thin microtubule (MT)-based membrane protrusions known as microtentacles (McTNs), as shown for mammary epithelial cells \cite{Whipple07}. While McTNs promote tumor cell reattachment to other cells, extracellular matrix or endothelial monolayers \cite{Balzer10,
Boggs15,Matrone10,Whipple07,Whipple10}, only indirect evidence exists of McTNs facilitating adhesion to blood vessels \cite{Korb04}. Thus, little is known about the mechanism underlying tumor cell adhesion to vessel walls and the role McTNs play in this process. In general, membrane protrusions play key roles in processes such as cell migration  \cite{Bouchet17,
EtienneManneville13,Yamada19,Caswell18}, signal conduction \cite{Laughlin03,
Buszczak16}, and environmental sensing \cite{Mattila08,Marshall06}. Most protrusions are actin- \cite{Clainche08,Caswell18,Linder23} or MT-based \cite{Bender15,Lu13,Whipple07} and the required forces to deform the membrane are generated by either polymerization dynamics  \cite{Peskin93,
Theriot00} or sliding of filaments with respect to each other powered by molecular motors  \cite{Jakobs15,Oelz18}. Polymerization of actin filaments contributes to the development of diverse types of cell protrusions including lamellipodia, filopodia, and invadosomes \cite{Clainche08,Linder23}. Sliding of MTs by means of molecular motors facilitates processes such as neurite outgrowth \cite{Lu13}, axonal regeneration \cite{Lu15}, and formation of proplatelets \cite{Bender15}. Previous studies suggest that MT polymerization can also contribute to protrusion formation  \cite{Fygenson97,Emsellem98,
Gavriljuk21,Bouchet16}. To gain insight into how McTNs formation in CTCs can be controlled, prevented or reversed, a detailed understanding of the mechanism driving McTN formation is crucial. Here we define the dominant mechanism of McTNs formation and suggest a mechanism how these long, curved protrusions may enhance cell adhesion to vessel walls.

\section{Materials and Methods}
\subsection{Cell culture}
Immortalized retinal pigmented epithelium (hTERT-RPE1) cells were cultured in Dulbecco's modified Eagle's medium/F12 (Gibco) with 10 $\%$ fetal bovine serum (Fisher Scientific), 1$\%$ GlutaMAX (Fisher Scientific), and 1$\%$ penicillin/streptomycin (Gibco) at 37 °C and 5$\%$ CO$_2$.  MT dynamics were studied  using RPE-1 cells that stably expressed mEmerald-vimentin and TagRFP-tubulin, which were kindly provided by Gaudenz Danuser (UT Southwestern, Dallas, TX, USA).  EB3 experiments were performed using RPE-1 cells that were transiently transfected with an EB3-tdTomato plasmid. The plasmid was kindly provided by Kristian Franze (Institute for Medical Physics, Friedrich-Alexander-Universität Erlangen-Nürnberg, Erlangen, Germany). Transfection was performed using X-tremeGENE$^{tm}$ 9 DNA (Sigma) as the transfection reagent, and 200 $\frac{\mu g}{ml}$ G418 (Sigma) was employed for selection purposes. Following a two-week selection period, over 90$\%$  of cells displayed fluorescence, which persisted for several weeks.

\subsection{Generation of McTNs}
To generate McTNs in RPE-1 cells, cells were detached from the surface using trypsin, centrifuged at 1300 rpm for 3 minutes, and then resuspended in media containing 1 $\mu$M latrunculin A (Sigma-Aldrich) and 5 mM Hepes (Gibco). This treatment leads to formation of McTNs within 10 minutes and is reversible by washout of latrunclin A. To prevent displacement of McTNs during imaging, glass-bottomed microscopy dishes (FluoroDish Cell Culture Dish, FD35-100, WPI Europe) were prepared. The dishes were coated with Cell Tak (Corning® Cell-Tak(TM) Cell and Tissue Adhesive, Merck), which was diluted in 0.1 M sodium bicarbonate. Each sample was imaged for a maximum duration of 1 hour.

\subsection{Cell adhesion measurement}
The experiment was carried out using either untreated RPE-1 cells or RPE-1 cells treated with 1 $\mu$M latrunculin A to induce McTN formation. Throughout all incubation periods, cells were maintained at 37 °C with 5$\%$ CO$_2$.
Suspended RPE-1 cells (100,000 cells per well) were placed into each well. Medium was changed at intervals of 20, 40, 60, or 90 minutes to remove unattached cells. After 90 minutes, cells were incubated for an additional 4 hours. 
The number of attached cells was determined using an MTT assay (Sigma). The cell medium was replaced with an MTT-medium solution at a concentration of 0.5 mg/ml and incubated for 40 minutes. Following this incubation, the MTT solution was replaced with DMSO to dissolve the MTT crystals. The cell number correlates with the intensity of the MTT color, which was measured using a Tecan plate reader. Each well was measured three times, and the mean value was taken.  Further, 3-4 samples per condition were measured and the experiment was repeated on 3 different days. All values were normalized to the intensity value of a reference for the respective condition well in which cells adhered for 5 hours and 30 minutes. Due to small variance in the exact initial cell number and the number of attached cells, relative values bigger than 1 are possible.

\subsection{Microscopy}
All experiments were performed using a spinning disk unit (CSU W1; Yokogawa, Andor Technology, Belfast, UK) with a pinhole size of 25 $\mu$m, coupled to an inverted microscope (Eclipse Ti-E; Nikon, Tokyo, Japan). Images were recorded using a digital camera with a 6.5 $\mu$m pixel size (flash 4.0; Hamamatsu, Hamamatsu City, Japan). 
Samples were visualized with an oil immersion objective with 60 x magnification and a numerical aperture of 1.4. A temperature of 37 °C was maintained throughout the experiments using a heating unit (Okolab).  25 mM HEPES (Gibco) was added to cell culture medium during the experiment to stabilize the pH-value.

\subsection{Image processing} We developed a C{\footnotesize{++}}-based package for 
image processing which enabled extraction of structural characteristics such as length and curvature of the 
McTNs. The procedure started with reconstruction of the three-dimensional shape 
of the cells from the successive images taken at different heights. The three-dimensional 
cell shape is required to correctly track the variations in the curvature and thickness 
of each McTN and to monitor McTN dynamics and assign an accurate 
length to them. By combining various existing image processing approaches, we developed 
a new framework for distinguishing the McTNs from the cell body. By calculating 
the directional density correlations from the pixel intensity information, we found that 
the anisotropic decay of the correlations differs greatly between the McTNs
and the cell body; as a result, it is possible to decompose the cells into the main 
cell body and the McTNs. In addition to determining the cell-body boundary 
and the outline of the entire cells, assigning a length to a curved tube with varying 
thickness is also a challenging task. The details of our image processing technique 
will be published elsewhere. Using the package developed, we can: (i) estimate the 
McTN  length by finding the shortest path along the protrusion body to reach 
the McTN tip from the nearest cell-body boundary point; (ii) estimate the 
number of McTNs and their total lengths; (iii) quantify other morphological 
properties of interest, including variations in curvature and thickness of the 
McTN; and (iv) track the dynamics of the McTNs by following the 
time-lapse images, {\it e.g.}, during the growth of the McTNs.

\subsection{EB3 experiments}
To generate McTNs, cells were prepared as described in {\it Generation of McTNs}. Time-lapse images were captured from various cells, with the time intervals and overall duration varying across experiments. Experiments were performed on three different days. For analysis, the McTN of interest was traced in Fiji by drawing a segmented line along the McTN. The line originated from the onset of the McTN at the cell body and extended towards the tip of the McTN. In cases where the distance of the McTN tip changed during the time lapse, the maximum distance was recorded. The line thickness was set to 9 pixels, which corresponds to an average McTN thickness of 0.99 $\mu$m.
A kymograph was generated along this line, representing the mean intensity in the transverse direction of the line.
The kymographs were manually analyzed to determine the frequency of MTs entering the McTNs, as well as the frequency of rescue and catastrophe events. To accomplish this, we counted the occurrences of EB3 tips appearing from the McTN base, appearing along the McTN, or disappearing. These counts were then divided by the duration of the experiment and the estimated number of MTs within the McTNs, which we approximated as 10.
To analyse the velocities, we used the Kymobutler software which detects tracks within kymographs and calculates several parameters such as the velocities \cite{Jakobs19}. The shown velocities are the mean frame to frame velocities.

\subsection{FRAP experiment}
To generate McTNs, the cells were prepared following the previously described method. To modify the relative contribution of MT sliding, cells were additionally treated with either 5  $\mu$M paclitaxel (Sigma) or 100 $\mu$M  Kinesore (Sigma).
For bleaching, a laser with a wavelength of 561 nm and 70$\%$ of maximum power (70 mW/mm$^2$) was used. The bleaching area consisted of a 1 $\mu$m x 1 $\mu$m square and was manually located approximately at half the length of the McTN length. During image acquisition, a laser with a wavelength of 561 nm and 25$\%$ of the maximum power was employed, with an exposure time of 200 ms. The videos consistently followed the same sequence, with images taken at intervals of 500 ms. Image acquisition began 5 s prior to the bleaching pulse, which lasted for 100 ms. Subsequently, images were acquired for a duration of 600 s. Throughout the experiment, the Nikon Perfect Focus System was enabled to prevent focus drift. For each condition 23-24 videos were acquired over the course of different days.
The videos were analyzed using a custom MATLAB script following the protocol outlined by Fritzsche and colleagues \cite{Fritzsche15}. In summary, the intensity values were corrected for background signal and unintentional photobleaching. Square regions of interest (ROIs) were manually selected for ROI1, ROI*, and ROI2 to cover 1 $\mu$m of the McTN (see Suppl. Fig. S4). The mean intensity within these regions was measured over time.
To normalize the intensities, the values in all three ROIs were divided by the respective mean intensity within the first 5 seconds before the bleaching pulse. In ROI*, the intensity immediately after the bleaching pulse was set to zero as a reference point.
Bar plots were generated by first calculating the mean intensity across all experiments within a specific condition. Subsequently, the mean intensities were determined for the following time intervals: t = [0, 60s], t = [0, 300s], and t = [0, 600s]. These choices enabled us to compare the cumulative changes of intensity after short, intermediate, and long times.

\subsection{Statistical analysis and graphical representation}
All experimental graphs were generated using the Seaborn library in Python 3.
In the bar plots, the box represents the mean value, while the error bars indicate the standard deviation.
For the boxplots, the line within the box represents the median value, and outliers are excluded from the visualization.
Statistical analyses were conducted using the SciPy library in Python 3. A T-test (ttest\_ind) was applied unless the data did not meet the assumption of normality, in which case a Mann-Whitney U-Test (mannwhitneyu) was performed. Data were checked for normality by conducting the Shapiro-Wilk-Test in Python 3. 
The p-values for all statistical tests are provided in the supplemental information.

\subsection{Modeling MT growth against soft barriers}
We considered a two-state model for MT dynamics and growth against a soft barrier \cite{Dogterom93,Ebbinghaus11,Chakraborty07}, with MT filaments being linear rigid rods consisting of subunits $\delta\,{\simeq}
\,0.6\,$nm. The MT tip dynamics was described by four parameters: growth 
$r\!_{_G}$ and shrinkage $r\!_{_S}$ rates and catastrophe $\nu\!_{_C}$ and rescue 
$\nu\!_{_R}$ frequencies. All parameters were extracted from our experimental EB3 and MT kymographs and are summarized in Table S1. By dividing the velocity by the length of one subunit $\delta\,{\simeq}$ \,$0.6\,$nm and taking the inverse, we obtain the corresponding rates.
\begin{table}[h]
\centering
\caption{Parameters describing MT dynamics}
\begin{tabular}{l*{1}{l}c} 
\hline
Parameter&Value \\ 
\hline
Force-free polymerization velocity $v\!_{_{G}}^{\,_0}$ &  0.25$\, \mu  \text{m}$ $ \text{s}^{-1}$ \\
Depolymerization velocity $v\!_{_{S}}^{\,_0}$ &  0.4$\, \mu  \text{m}$ $ \text{s}^{-1}$ \\
Force-free catastrophe frequency $\nu\!_{_{C}}^{\,_0}$ &  0.004$\,\text{s}^{-1}$ \\
Rescue frequency $\nu\!_{_R}$ &  0.0005$\,\text{s}^{-1}$ \\
\hline
\end{tabular}
\end{table}
We used a piecewise linear 
deformation-force model depicted in Fig. Supp. 5A, where the force linearly 
grows with the membrane displacement for small deformations compared to the cell 
size, but eventually reaches an asymptotic level $\mathcal{F}$ for longer 
deformations \cite{Derenyi02,Powers02,Sheetz01}. According to the Brownian ratchet model \cite{Peskin93,Dogterom02,
Stukalin04}, the polymerization rate decays with the force exerted on the tip of 
a MT of length $d$ as
\begin{equation}
r\!_{_G}(d)=r\!_{_{G}}^{\,_0}\;\exp\Big[{-}\frac{\delta}{k\!_{_B}
T}\,F(d)\Big],
\label{Eq:rG}
\end{equation}
where $r\!_{_{G}}^{\;_0}$ is the polymerization rate of a freely growing MT, 
$k\!_{_B}$ the Boltzmann constant, and $T$ the temperature. The asymptotic 
polymerization rate is given by $r\!_{_{G}}^{\,_\infty}=r\!_{_{G}}^{\,_0}
\,exp[{-}\mathcal{F}\delta{/}k\!_{_B}T]$. As well as limiting the rate of 
tubulin addition, the force also reduces the mean time interval $\tau\!_{_C}$ 
between successive catastrophe events, as 
$\tau\!_{_{C}}(d)\!=\tau\!_{_{C}}^{\,_0}{-}\displaystyle\frac{r
\!_{_{G}}^{\,_0}{-}r\!_{_G}(d)}{r\!_{_{G}}^{\,_0}{-}r\!_{_{G}}^{\,_\infty}}(\tau
\!_{_{C}}^{\,_0}{-}\tau\!_{_{C}}^{\,_\infty}\!)$, where $\tau\!_{_{C}}^{\,_0}$ 
and $\tau\!_{_{C}}^{\,_\infty}$ are the force-free and asymptotic time intervals 
between the catastrophe events, respectively \cite{Janson03}. Thus, the catastrophe 
frequency grows as 
\begin{equation}
\nu\!_{_{C}}(d) = \frac{\nu\!_{_{C}}^{\,_0}}{1{-}\displaystyle\frac{r
\!_{_{G}}^{\,_0}{-}r\!_{_G}(d)}{r\!_{_{G}}^{\,_0}{-}r\!_{_{G}}^{\,_\infty}}
\Big(1{-}\frac{\nu\!_{_{C}}^{\,_0}}{\nu\!_{_{C}}^{\,_\infty}}\Big)},
\label{Eq:vC}
\end{equation}
where $\nu\!_{_{C}}^{\,_0}{=}1{/}\tau\!_{_{C}}^{\,_0}$ and $\nu
\!_{_{C}}^{\,_\infty}{=}1{/}\tau\!_{_{C}}^{\,_\infty}$.

We also introduce $p^{_+}\!(d,t)$ and $p^{_-}\!(d,t)$ as the probabilities of 
having a filament of length $d$ at time $t$ in the growing or shrinking state, 
respectively. The evolution of these probabilities can be described by the 
following master equations:
\begin{eqnarray}
\left\{ \begin{array}{ll}
\!\!\dot p^{_+}\!(d,t)=r\!_{_G}(d{-}1)\;p^{_+}\!(d{-}1,t)-r\!_{_G}(d)\;p^{_+}
\!(d,t)\vspace{1mm}\\
\hspace{13.2mm}+\nu\!_{_R}\;p^{_-}\!(d,t)-\nu\!_{_{C}}(d)\;p^{_+}\!(d,t),\vspace{2.5mm}\\
\!\!\dot p^{_-}\!(d,t)=r\!_{_S}\;p^{_-}\!(d{+}1,t)-r\!_{_S}\;p^{_-}\!(d,t)+
\nu\!_{_C}(d)\;p^{_+}\!(d,t)\vspace{1mm}\\
\hspace{13.2mm}-\nu\!_{_{R}}\;p^{_-}\!(d,t). 
\end{array}
\right.
\label{Eq:MasterEqs}
\end{eqnarray}
To demonstrate the evolution of the MT length, these equations were solved by 
considering the appropriate boundary conditions. For example, a nucleation rate, 
or alternatively, an instantaneous rescue event, can be defined to prevent shrinkage 
below $d{=}0$ ({\it i.e.}, reflecting boundary conditions); in the latter case, the 
following constraints were imposed
\begin{eqnarray}
\left\{ \begin{array}{ll}
\!\!\dot p^{_+}\!(0,t)=r\!_{_S}\;p^{_-}\!(1,t)-r\!_{_G}(0)\;p^{_+}\!(0,t),
\vspace{2mm}\hspace{20mm}\\
\!\!p^{_-}\!(0,t)=0.\hspace{20mm}
\end{array}
\right.
\label{Eq:BCLzero}
\end{eqnarray}

Similarly, to take into account the limited stretchability of the cell surface, 
an upper filament length threshold can be imposed by an instantaneous catastrophe 
event at $d{=}d_{max}$, which leads to 
\begin{eqnarray}
\left\{ \begin{array}{ll}
\!\!\dot p^{_-}\!(d_{max},t){=}r\!_{_G}(d_{max}{-}1)\,p^{_+}\!(d_{
max}{-}1,t){-}r\!_{_S}\;p^{_-}\!(d_{max},t),
\vspace{2mm}\\
\!\!p^{_+}\!(d_{max},t){=}0.
\end{array}
\right.
\label{Eq:BCLmax}
\end{eqnarray}

By setting all time derivatives to zero, the probability distribution of the filament 
length in the steady-state can be deduced \cite{Govindan04}. We numerically obtained the stationary 
distribution of the microtubule length $p(d)=p^{_+}\!(d)+p^{_-}\!(d)$ and, thus, 
the probability distribution of the protrusion length $P(\ell)$ in terms of the 
asymptotic force $\mathcal{F}$.

\subsection{Modeling semiflexible filaments}
To model semiflexible McTNs, we considered a filament of total length $L$ divided 
into $n{+}1$ segments with length $l{=}\frac{L}{n{+}1}$. Thus, the centerline of 
the filament was segmented into $n{+}1$ edges $\mathbf{x}_{_0}\!, \mathbf{x}_{_1}\!, 
{...}, \mathbf{x}_{_n}\!$. The elastic energy $E$ of the filament consists of 
bending $E_{\,bend}$ and extensional $E_{\,ext}$ energy terms \cite{Weber18}. 
Using $\mathbf{e}_i\,{=}\,\mathbf{x}_{i{+}1}{-}\,\mathbf{x}_i$ and 
introducing curvature $\kappa_i\,{=}\,2\tan\frac{\phi_i}{2}$ ($\phi_i$ being 
the turning angle between two consecutive edges), the bending energy term is 
given by \cite{Kratky49}
\begin{equation}
E_{\,bend}\,{=}\,Y I \sum_{i=1}^{n{-}2}\frac{(\kappa\mathbf{b})_i^2}{l},
\end{equation}
with $Y$ being the Young's modulus, $I$ the moment of inertia of each segment, 
and $(\kappa\mathbf{b})_i$ the discrete curvature binormal defined as $(\kappa
\mathbf{b})_i\,{=}\,\frac{2\,\mathbf{e}_{i{-}1}{\times}\mathbf{e}_i}{l^2{+}\,
\mathbf{e}_{i{-}1}{\cdot}\,\mathbf{e}_i}$. To conserve the overall length of 
the filament, we employed an extensional energy term as \cite{Camalet00}
\begin{equation}
E_{\,ext}\,{=}\,\frac{Y\!A\,l}{2}\sum_{i{=}0}^{n{-}2}\!\left(\frac{|\!|
\mathbf{e}_i|\!|}{l}{-}1\right)^2\!\!\!,
\end{equation} 
with $A$ being the cross section of the filament. The total energy of the filament 
was obtained by adding the contribution of the adhesion energy (introduced in 
the main text) to the bending and extensional energies. Finally, by minimizing 
the total energy, the optimal configuration of the filament was obtained.

\section{Results}
\subsection{Generation of McTNs in non-cancerous RPE-1 cells}
To determine the driving mechanism of McTN formation, specifically whether 
McTNs arise from microtubule (MT) polymerization or sliding (Fig.\ 1A), we first identified a sufficient model cell line. We hypothesized that the mechanism of McTN formation is independent of the cancerous phenotype of the cell, as sufficient weakening of the actin cortex of any cell is the major underlying mechanism \cite{Matrone10b}.Previously, McTN formation was found to be induced or enhanced by destabilization of the actin cortex with drugs such as latrunculin A \cite{Whipple07}. Thus, latrunculin A treatment of suspended RPE-1 cells, which endogenously express TagRFP-tubulin, results in the formation of membrane protrusions filled with MTs (Fig.\ 1E,  more exemplary images of RPE-1 cells in Suppl. Fig.\ S1A). Moreover, these protrusions phenomenologically resemble McTNs shown in literature (Fig.\ 1B-D). These latrunculin A-induced membrane protrusions grow within 15 minutes, as reported by Killilea et al. \cite{Killilea17}, and persist for hours.  
Three random RPE-1 cells were selected at 15 minutes after latrunculin A treatment and the length of their three-dimensional protrusion was measured over a period of 80 to 120 minutes. For each cell and each time point the mean length of all protrusions was determined. To compare the length dynamics of these three cells, we normalized the protrusion length by dividing by the mean length during the first 60 minutes. Our data show that, despite minor fluctuations, the latrunculin A-induced protrusions continue to grow for roughly an hour after their initial formation (in the first 10 minutes), and then reach a steady state (Fig. 1F). Consequently, although these protrusions consist of a MT bundle, they do not exhibit the dynamic instability typically expected of individual MTs. Since McTNs promote the reattachment of cells \cite{Whipple07},  we checked whether latrunculin A-induced protrusions formed by RPE-1 cells fulfil this same function. Treated or untreated RPE-1 cells were plated in suspension. To quantify cell attachment within a given time, cell media were exchanged after 20, 40, 60, and 90 min. Cells that had already attached remained in the dish and were counted. At shorter time points, a higher amount of treated RPE-1 cells with MT-based protrusions attached to the surface than did untreated cells. The relative difference of attached cells decreased with time but was still clearly visible after 90 min (Fig. 1G). This clearly shows that latrunculin A-induced protrusions in RPE-1 cells promote reattachment of these cells. Furthermore, as depolymerization of actin enhances adhesion, the mechanism appears to be independent of f-actin  \cite{Korb04}. Experiments were then repeated on surfaces with different coatings:  Pll-g-PEG-coating produces a non-adhesive surface and fibronectin coating enhances focal adhesion and thus actin-based adhesion. While untreated RPE-1 cells clearly adhere best on a fibronectin-coated surface, adhesion of cells with latrunculin A-induced protrusions was independent of the coating (Suppl. Fig. S1B). This confirms that the adhesion mechanism of RPE-1 cells with latrunculin A-induced protrusions is independent of actin. Further, it shows that these cells can even adhere on a non-adhesive surface presumably because these thin protrusions attach to micrometer sized defects in the Pll-g-PEG coating. 
Due to the similarity of latrunuculin A-induced protrusions to McTNs and their ability to promote adhesion in an f-actin independent mechanism, we conclude that the protrusions formed by RPE-1 cells are McTNs. This further demonstrates that the mechanism of McTN formation is valid beyond the context of cancer cells.  We therefore continued our study using RPE-1 cells.

\begin{figure}[htbp]
\centering
\includegraphics[width=0.8\linewidth]{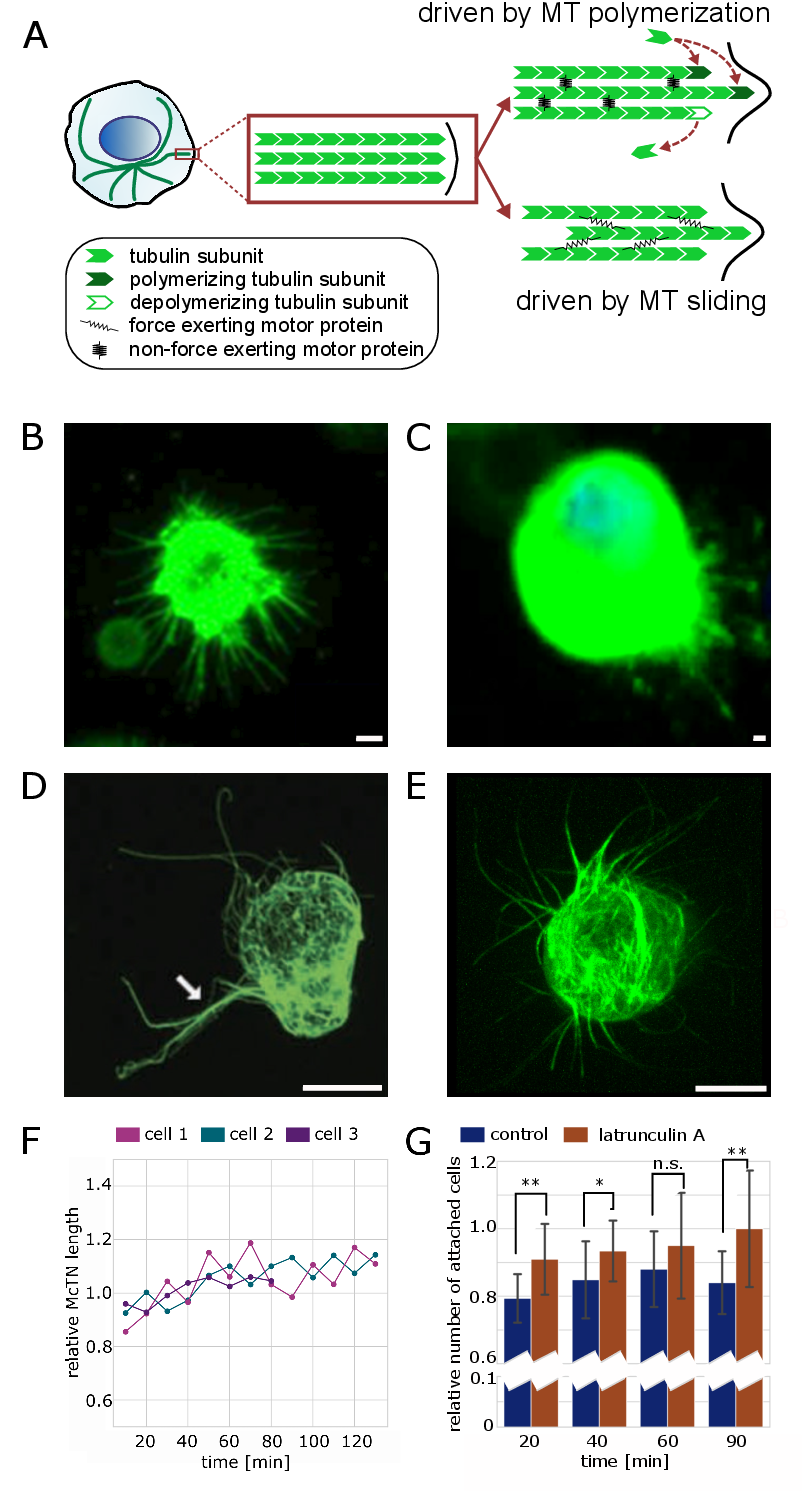}
\caption{{\bf Microtentacles in CTCs and RPE-1 cells. } (A) Schematic representation of possible mechanisms of MT-based protrusion formation at the cell membrane. This process can be driven by MT polymerization dynamics or by sliding of MTs against each other powered by molecular motors. (B) - (E) McTNs have different shapes, numbers, and lengths depending on the cellular system (Scale bars:~10 $\mu$m), (B) McTNs in the CTC cell line CTC-MCC-41. Membrane stained with WGA. Image taken from \cite{Vardas23}. (C) McTNs formed by a CTC of a TNBC patient. Membrane stained with WGA. Image taken from \cite{Vardas23}. (D) McTNs formed by a cancer HCC1428 cell. Alpha tubulin staining. Image taken from \cite{Matrone10}. (E) McTNs formed by a non-cancer RPE-1 cell, alpha tubulin staining. (F) McTN length over a time course of 80 - 120 minutes of three randomly chosen cells. For each cell and each time point, mean length of all McTNs was determined. Values were normalized by the mean length measured during the first 60 minutes. (G) McTNs improve cell attachment. The relative number of attached cells increases over time of McTN generation. Data collected at 20, 40,  60, and 90 min. Untreated cells (blue bars) are compared to cells treated with 1 $\mu$M latrunculin A (red bars) to induce McTN formation. The relative number of attached cells was determined with MTT assay. Attached cells after 5h 30 min serve as a reference value ($^{***}$p${\leq}0.001$,  $^{**}$p${\leq}0.01$,  $^{*}$p${\leq}0.05$, n.s., not significant; Mann-Whitney U-Test, n ${\geq}$6).}
\label{Fig1}
\end{figure}

\subsection{McTNs consist of a bundle of parallel MTs anchored in the MT organizing center (MTOC)} 
Since McTNs are MT-based membrane protrusions, we asked whether the organization of MTs in McTNs resembles that in other MT-based protrusions. In neurites or proplatlets, for example, MTs exist in fragmented or bundled states with mixed polarities \cite{Kapitein15, Hartwig03, Italiano99}. When adhered to a substrate, RPE-1 cells have a clearly identifiable MTOC from which the majority of MTs emerge (Suppl. Fig. S2A). However, detachment from substrate could alter the organization of MTs. Due to the high MT density in suspended cells, it is challenging to identify individual MTs inside the cell and trace them back to the MTOC. Tracking of individual plus-end binding proteins EB3 from the MTOC to the McTN tip was also not possible due to the high MT density.  Instead, we looked at the orientation of MTs by studying their polymerization dynamics inside McTNs. MTs display dynamic instability, which means that they constantly change between a growing and a shrinking state. Changing from the growing to the shrinking state is termed {\it catastrophe} and switching from the shrinking to the growing state is 
termed {\it rescue} \cite{Dogterom93}.To study MT polymerization dynamics inside McTNs, RPE-1 cells were transfected with a plasmid for EB3-RFP, a fluorescent protein that binds to the growing plus-ends of MTs. All experiments were performed 15 minutes after latrunculin A treatment, and a sample was imaged for a maximum of one hour to have McTNs in a stable or slightly growing phase (see Fig. 1F). Time-lapse images of cells with observable EB3 signal within McTNs were acquired (Suppl. Movie S1) and kymographs were generated (Fig. 2A). From the kymographs, we quantified the appearance and disappearance of EB3 signals and, based on \cite{Killilea17}, estimated an average of 10 MTs per McTNs. Consequently, we determined the frequency of new MTs entering McTNs and the rescue frequency (Fig. 2B) and as a result, determined that most EB3 signals originated from within the cell body. Furthermore, almost all EB3 tracks indicated that MTs polymerize from within the cell body towards the McTN tip (anterograde). Only rarely did EB3 tracks move towards the cell body (retrograde) (Fig. 2C). To confirm that this is a result of retrograde polymerization, we compared the velocities of anterograde and retrograde moving EB3 tracks and found no significant difference (Fig. 2D). We assume that retrograde polymerization events result from spontaneous nucleation events which lead to MTs with opposite orientation. Nevertheless, such nucleation events are very rare, and the frequency of retrograde growth is very small. Because of this and since most MTs emerge from within the cell body, the existence of fragmented MTs inside McTNs without connection to the MTOC is unlikely. This is further supported by scanning electron images which only show continuous MTs emerging from within the cell body (Suppl. Fig. S3). We conclude that inside McTNs the majority of MTs have their plus ends facing the McTN tip, that they originate from the cell body, and that they are likely connected to the MTOC. Super resolution imaging by Killilea and colleagues led to the same conclusion \cite{Killilea17},. Moreover, such connection to the MTOC potentially serves as a stabilizing point which helps MTs to generate the pushing forces required for protrusion of the cell membrane.

\begin{figure}[htbp]
\centering
\includegraphics[width=0.99\linewidth]{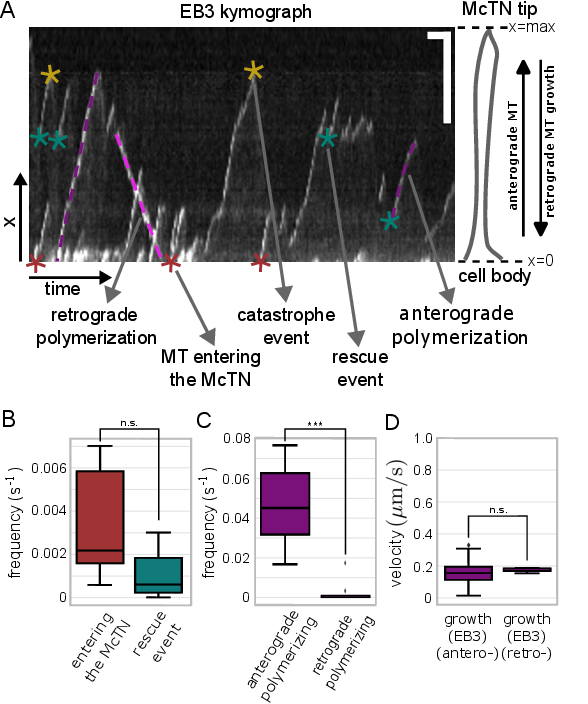}
\caption{{\bf MT dynamics within McTNs.} (A) Exemplary kymograph of MT plus-end 
binding protein EB3 within a McTN. Different colors mark examples of different 
dynamic events, such as anterograde (dark purple lines) and retrograde (light 
magenta line) polymerization, rescue (blue stars) and catastrophe (yellow stars) 
events, and new MTs entering the McTN from the cell body (red stars). Scale bars: 
time, 60 s; space, 5 $\mu$m. (B) Frequencies of MTs entering the McTN and rescue 
events ($N{=}$16). (C) Frequencies of anterograde and retrograde polymerizing EB3 
comets ($N{=}$16). (D) Anterograde and retrograde polymerization velocities from 
EB3 kymographs ($N{=}$35). ***p$\leq$0.001; n.s., not significant 
(Mann-Whitney U-Test).}
\label{Fig2}
\end{figure}

\begin{figure*}[htbp]
\centering
\includegraphics[width=0.99\linewidth]{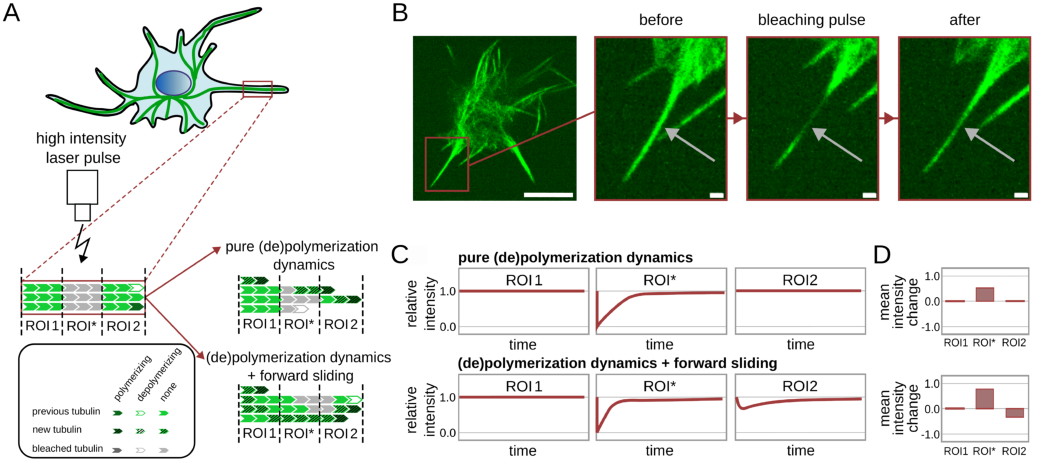}
\caption{{\bf Fluorescence intensity recovery after photobleaching (FRAP) experiments.} 
(A) Schematic representation of McTN FRAP experiment. A high intensity laser pulse bleaches the fluorescent MT bundle in a region of interest (ROI*). Two possible scenarios for changes in fluorescence intensity of two adjacent regions (ROI1 and ROI2) are depicted. For pure MT polymerization dynamics, the bleached tubulin subunits cannot enter the ROI1 and ROI2 regions, thus, the intensities of these regions remain unchanged. Contrarily, additional forward sliding dynamics of MTs can push bleached tubulin subunits into ROI2 and reduce intensity. (B) Exemplary images of a bleached region (indicated with arrows) over the course of the experiment. Before applying the bleaching pulse on a selected McTN (marked in the left frame), the entire MT bundle is fluorescent. The bleaching pulse results in a local loss and subsequent recovery of fluorescence over time. Scale bars: 10 $\mu$m (left panel), 1 $\mu$m  (rest of panels). (C) Expected time evolution of fluorescence intensity in ROI*, ROI1, and ROI2 regions for two possible scenarios. (D) Expected mean intensity change within a given time interval with respect to the intensity at the bleaching time point for both scenarios and three different regions of interest.}
\label{Fig3}
\end{figure*}

\subsection{Visualizing MT sliding dynamics through fluorescence recovery after photobleaching (FRAP) experiments}
As shown above, MTs inside McTNs polymerize; however, an additional sliding mechanism may underlie McTN formation and could be linked to the functional role of McTNs. To address this, we performed fluorescence recovery after photobleaching (FRAP) experiments. A high-intensity laser pulse was directed at a region of interest (ROI*) along the stable McTN, leading to fluorescence loss in the defined ROI* and subsequent intensity recovery (Fig. 3B, Suppl. Movie S2). ROI* was manually located at approximately the middle of the McTN. The mean intensity within the region was calculated and all values were normalized against the initial mean intensity value (for more details see {\it Materials and Methods}). Due to the observed fast polymerization dynamics of MTs inside McTNs, the recently introduced mathematical approach to study filament transport \cite{Dallon22} cannot be applied here. Instead, intensity data were extracted from the ROI* as well as from two adjacent regions:  ROI1 (towards the cell body) and ROI2 (towards the McTN tip) (Fig. 3A). All three regions (ROI*, ROI1, and ROI2) were manually selected to cover approximately 1 $\mu$m of the McTN (exemplary images in Suppl. Fig. S4). Typically, for FRAP experiments intensity-over-time curves are presented (compare to Fig. 3C). However, for better visualization of different scenarios, we introduce an alternative representation in which the mean intensity was calculated within a time interval, T, and the intensity at the bleaching time point was subtracted (for more details see {\it Materials and Methods}). This yields a dimensionless quantity which we termed {\it mean intensity change} (Fig. 3D). For scenarios of pure MT polymerization dynamics inside McTNs, we anticipate fluorescence recovery in ROI* due to i) new MT polymerization into ROI*, ii) MT depolymerization followed by rescue events in front of ROI* (relative to the cell body), or iii) MT depolymerization with rescue events within ROI* (Fig. 3A). In these scenarios, the mean intensities in ROI1 and ROI2 remain unchanged (Fig. 3C, D). Next, we consider the presence of additional sliding dynamics. While it should be noted that retrograde and bidirectional sliding may also occur, to investigate the basic concept we assessed forward (anterograde) sliding. In a mixed polymerization and forward sliding scenario, the fluorescence intensity in ROI2 (or ROI1 in the case of backward sliding) should also be affected. While the recovery of fluorescence intensity in ROI* is accelerated due to fluorescent MTs sliding from ROI1 into ROI*, the intensity in ROI2 decreases because of bleached MTs sliding into ROI2 (Fig. 3A, C, D). The relative contribution of sliding and polymerization mechanisms dictates the magnitude of difference between the increase of intensity in ROI* and ROI2. In the extreme case of pure sliding, the increase of intensity in ROI* would equal the decrease of intensity in ROI2. Note, that diffusion of free tubulin can also change the fluorescent intensity. However, the highly confined geometry of McTNs leads to drastically reduced diffusion coefficients compared to diffusion in cytoplasm and is therefore insufficient to affect the observed trends.

\begin{figure*}[htbp]
\centering
\includegraphics[width=0.99\linewidth]{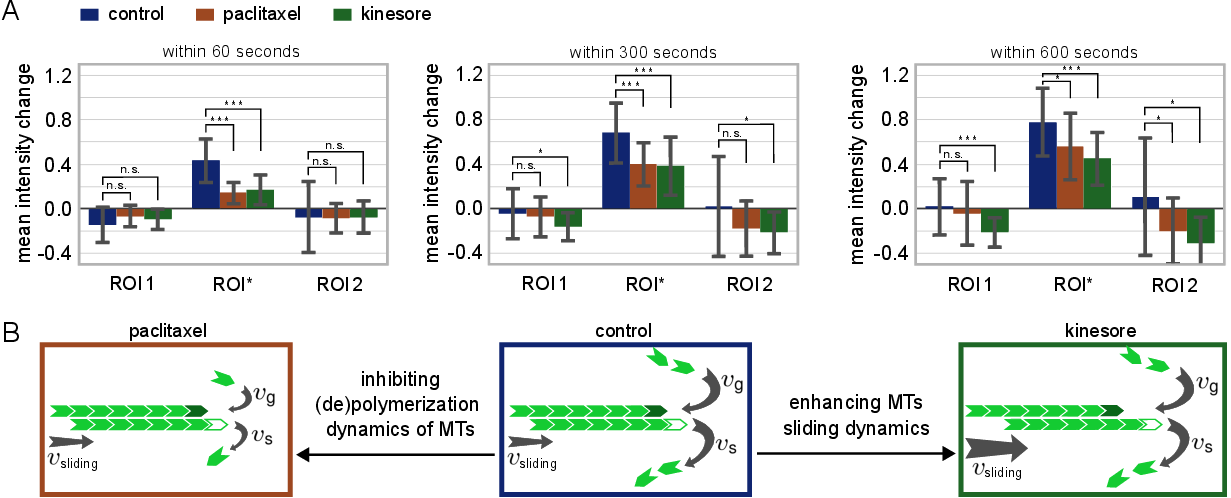}
\caption{{\bf Mean fluorescence intensity change after photobleaching under 
different conditions.} (A) Mean intensity change within first 60, 300, or 600 
seconds with respect to the intensity at the bleaching time in ROI1, ROI*, 
and ROI2. All cells were treated with 1 $\mu$M latrunuclin\,A. Additionally, 
5 $\mu$M paclitaxel or 100 $\mu$M kinesore was used to inhibit MT polymerization 
or enhance MT sliding, respectively. Number of experiments:\ control: $N{=}\,$23 
(blue); paclitaxel:\ $N{=}\,$23 (brown); kinesore:\ $N{=}\,$24 (green).  Error bar shows standard deviation. (B) 
Schematic representation of the impact of paclitaxel and kinesore on MT dynamics. 
While paclitaxel inhibits MT polymerization, kinesore enhances the sliding 
dynamics of MTs.***p$\leq$0.001; **p$\leq$0.01; *p$\leq$0.05; n.s., not 
significant (t-tests).}
\label{Fig4}
\end{figure*}

\subsection{MT polymerization dominates over MT sliding dynamics inside McTNs}
We performed FRAP experiments using RPE-1 cells expressing TagRFP-tubulin.  To induce McTN formation, cells were treated with 1$\mu$M  latrunculin A. RPE-1 cells formed stable McTNs within 15 min. We refer to this condition as the {\it control} To explore the dynamics on different time scales, we determined mean intensity changes within time intervals of 60, 300, and 600 s. In control experiments, there is an initial fast recovery in ROI* (40$\%$ within 60 s) followed by a slower recovery (80$\%$ within 600 s) (Fig. 4A, Suppl. Fig. S5A). As explained above, sliding dynamics would manifest in an intensity reduction in one or both regions adjacent to ROI*. While a slight decrease in intensity occurred within the first 60 s in both regions, this effect diminished over time (Fig. 4A). Importantly, this intensity drop did not exceed the amplitude of intensity fluctuations (Suppl. Fig. S5A). We assume that such a slight reduction in intensity is potentially an artefact of the measurement (e.g. a transient enhancement of MT depolymerization caused by the bleaching pulse). Therefore, we interpret our control experiments such that polymerization of MTs has a dominant role. However, we cannot exclude a potential contribution of sliding obscured by the fast polymerization dynamics. To test this, we treated cells with 5 $\mu$M  of paclitaxel to inhibit MT polymerization \cite{Jolly10} (Fig. 4B). Our observations in EB3-transfected cells (Suppl. Movie 3) as well as the slower and reduced intensity recovery in ROI* (Fig. 4A) confirmed this inhibition. While treatment with paclitaxel had no clear influence on the intensity in ROI1, the intensity in ROI2 consistently decreased within 600 s after the bleaching pulse (Fig. 4A). These results demonstrate that there is an effect of forward sliding dynamics of MTs inside McTNs, likely due to kinesin-1 activity  \cite{Jolly10, Lu13}. To test this, cells were treated with 100 $\mu$M kinesore, a compound known to enhance kinesin-1 activity by binding to its cargo-binding domain, thus increasing its affinity for adjacent MTs \cite{Randall17} (Fig. 4B). If kinesin-1 mediates MT sliding dynamics within McTNs, kinesore treatment should increase the contribution of sliding dynamics relative to polymerization dynamics. In fact, the intensity in ROI2 did decrease substantially following kinesore treatment compared to the paclitaxel condition. Interestingly, ROI1 also exhibited a clear intensity decrease suggesting simultaneous backward sliding dynamics in kinesore-treated cells. Additionally, intensity recovery in ROI* also slowed down (Fig. 4A) possibly a result of i) backward sliding and ii) indirect effects of kinesore on MT polymerization dynamics. Taken together, we conclude that in control cells MT dynamics inside McTNs are primarily governed by polymerization and the effect of sliding only becomes visible when either MT dynamics or kinesin activity is changed. As further confirmation, we asked whether the additional contribution of sliding is essential for generating the forces required for McTN formation.

\subsection{MT polymerization drives McTN formation}
To disentangle the contributions of different MT dynamics on McTN formation, numerical simulations of MT dynamics and growth against a membrane were performed. The complex structure of MTs was simplified as rods consisting of subunits. The MTs were considered to be either in a growing (polymerization) or shrinking (depolymerization) state  \cite{Dogterom93,Ebbinghaus11,
Shaebani16,Shaebani20}. In such a two-state model, the dynamics at the MT tip is phenomenologically described by growth and shrinkage velocities as well as catastrophe and rescue frequencies. These parameters were extracted from the EB3 and MT kymographs (Fig. 2B, Suppl. Fig. S6B, C) (See {\it Materials and Methods} for master equations describing the time evolution of the MT length). MTs polymerize freely until they reach the cell membrane, which exerts a force on the MT tips. This force depends on properties of the membrane (e.g. bending rigidity and surface tension  \cite{Sheetz01}). Note that the force is non-monotonic for small deformation but reaches an asymptotic force $\mathcal{F}$  for deformations bigger than 1$\%$ \cite{Derenyi02,Powers02} of the cell radius, which McTNs are (Suppl. Fig. S7A). According to the Brownian ratchet model, the force exerted by the membrane affects the growth velocity \cite{Peskin93,Dogterom02,Stukalin04} and catastrophe rate \cite{Janson03} (see equations (1), (2) in {\it Materials and Methods}). Since McTNs consist of a bundle of MTs, the force is equally shared between MTs in close vicinity of the cell membrane \cite{Laan08,vanDoorn00,Krawczyk11}. Our FRAP experiments suggest that polymerization of MTs is the dominant mechanism for McTN growth. Therefore, we asked whether a pure polymerization mechanism could reproduce the length distribution of McTNs obtained in our experiments. The mean McTN length was calculated as a function of the force exerted by the membrane (Suppl. Fig. S7B). Comparison to the experiments shows that a force of $\mathcal{F}{\approx}\,$ 28.7 pN minimizes the error between simulated and experimental length distribution. Next, we checked whether we could also reproduce the FRAP experiments with these parameters. To repeat the FRAP experiment numerically, we marked all the MT subunits located in a region of length  1$\,\mu$m (ROI*) and extracted the time evolution of subunit exchange in this region. Our model with pure polymerization dynamics leads to a slower intensity recovery in ROI* as compared to the experiments (Fig. 5A). The simulation curve has an overall deviation of around 22$\%$ from the experimental data. We asked whether forward sliding would decrease the deviation from the experimental FRAP curves. Forward sliding dynamics were included in the simulations by employing a simple model:  MTs can randomly switch between sliding and non-sliding mode. As an approximation, the sliding velocity was set as equal to the MT growth velocity (see Suppl. Table S1). In sliding mode, the MT tips close to the McTN tip exert an additional force on the membrane. We varied the sliding force so that it reaches the best fit to the experimental data (Fig. 5A). Thus, with a sliding force of 3.2 pN the numerical recovery curve has an overall deviation of only around 5$\%$ from the experimental data. To further validate our model, we also reproduced paclitaxel and kinesore experiments, as well as the intensity evolution in all three ROIs. By reducing the MT polymerization dynamics or increasing the sliding force in FRAP simulations, the experimental results could be reproduced with nearly 10$\%$ of relative errors (Suppl. Fig. S7C). Eventually, by including the forward sliding dynamics into the model, the error of fit of McTN length distribution reached better values than allowed by a pure polymerization mechanism. The combination of MT polymerization and forward sliding generates McTNs with a mean length comparable to that of McTNs in the control experiments (4.1 $\pm$ 0.6 $\mu$m vs 4.3 $\pm$ 0.8 $\mu$m). Pure MT polymerization dynamics produce slightly shorter protrusions (3.9 $\pm$ 0.6 $\mu$m) (Fig. 5B). Thus, our simulations confirm that i) MT sliding dynamics play a role in McTNs, but the contribution is very weak and ii) polymerization is the dominant force-generating mechanism during formation of McTNs. Further, the force exerted due to polymerization is ten times higher than the force generated by sliding.

\begin{figure}[t]
\centering
\includegraphics[width=0.9\linewidth]{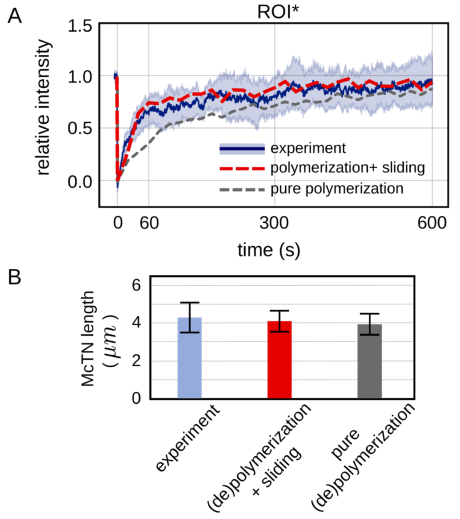}
\caption{{\bf MT growth and FRAP simulations}. (A) Time evolution of the relative 
intensity in ROI* from control experiments (blue line) and simulations with pure 
polymerization (gray) or polymerization plus forward sliding dynamics of MTs (red). 
(B) Mean McTN length obtained from control experiments, simulations with pure MT 
polymerization, or simulations with polymerization and sliding dynamics.}
\label{Fig5}
\end{figure}

\begin{figure*}[t]
\centering
\includegraphics[width=0.99\linewidth]{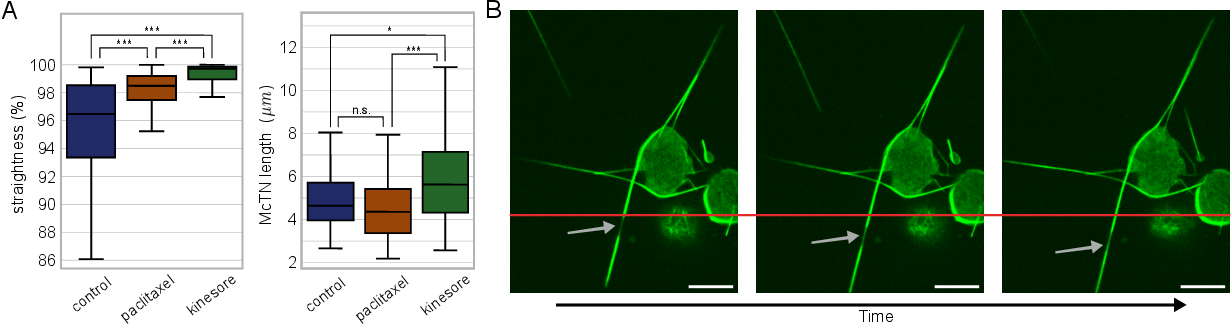}
\caption{{\bf Detachment of MTs from the MTOC by kinesore treatment.} (A) Straightness 
of McTNs defined as the end-to-end distance divided by the total length of the McTN 
(left) and mean length of McTNs (right) under different treatments ($^{***}$p${\leq}0.001$, 
n.s., not significant; Mann-Whitney U-Test, $n{\geq}100$). (B) Moving of a bleached 
area (marked with arrows) along the McTN due to sliding of MTs in a kinesore treated 
cell. The horizontal red line indicates the initial position of the bleached area. Scale bars:\ 10 $\mu$m.}
\label{Fig6}
\end{figure*}

\subsection{Kinesore detaches MTs from the MTOC and straightens McTNs}
The simulations of the FRAP study revealed that MT polymerization generates forces at the McTN tip that are an order of magnitude larger than forces due to MT sliding dynamics. Moreover, the presence of McTN tip forces and anchorage of MTs in the MTOC curves the McTN. Thus, MT polymerization also plays a key role in bending of McTNs, as is evident in the bending of McTNs in paclitaxel treatment experiments. The straightness of McTNs was quantified by dividing the Euclidean distance between the McTN base and tip by the actual length of the McTN. The reduction of MT polymerization dynamics in paclitaxel-treated cells decreases the McTN tip force, leading to straighter McTNs (Fig. 6A). The McTNs are also slightly shorter compared to the control cells due to reduced polymerization. In kinesore-treated cells, the contribution of sliding was significantly increased. Simulations revealed an almost four-fold higher sliding force ($\sim$11.4 pN) (Suppl. Fig. S7C). Thus, enhanced bending of McTNs should occur in kinesore-treated cells compared to control cells, assuming that MTs are anchored in the MTOC. However, Fig. 6A shows that the McTNs are even straighter following kinesore treatment. This suggests that MTs may have lost their anchorage in the MTOC. In fact, in approximately 20$\%$  of kinesore-treated cells the bleached ROI* was clearly displaced along the McTN (Fig. 6B). Such displacement over several microns would be restricted if MTs are anchored in the MTOC. Additionally, MT organization appeared to be changed in adhered kinesore-treated cells, as the MTOC, as the point MTs emerge from, was no longer visible. Instead, MTs accumulated at the cell periphery (Suppl. Fig. S2B). Thus, the loss of connection of MTs to the MTOC and a stronger sliding dynamic in kinesore-treated cells result in longer and straighter McTNs. The increased density of kinesin motors enhances the rescue rate \cite{Akhmanova22}  leading to longer MTs and, thus, longer McTNs. Moreover, a higher density of kinesin motors strengthens the crosslinking of MTs to each other in the bundle. This enhances the rigidity of the McTN. Since the force required to buckle a semiflexible filament grows with the square of its cross-section area, the forces acting on the McTN tip are insufficient to buckle or bend the McTN in kinesore-treated cells.

\begin{figure*}[t]
\centering
\includegraphics[width=0.99\linewidth]{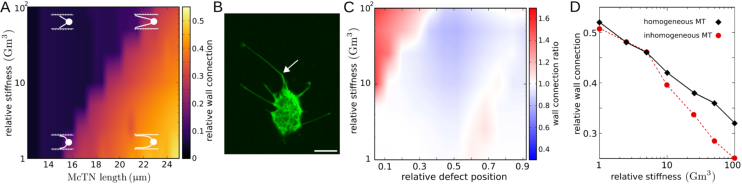}
\caption{{\bf McTN adhesion to vessel walls.} (A) Relative wall connection in terms of McTN length and relative stiffness of the McTN, $E{/}\epsilon$  (see text). (B) Example of a kink in a McTN of a kinesore treated cell indicated by white arrow. Scale bar:  10$\mu$m. (C) Ratio of the wall connection of a homogeneous McTN to that of a McTN with a kink in terms of the relative stiffness $E{/}\epsilon$  and the relative position of the kink with respect to the McTN tip. (D) Relative wall connection compared to the relative stiffness for filaments with homogeneous and inhomogeneous (i.e. with a kink) bending.}
\label{Fig7}
\end{figure*}

\begin{figure}[t]
\centering
\includegraphics[width=0.99\linewidth]{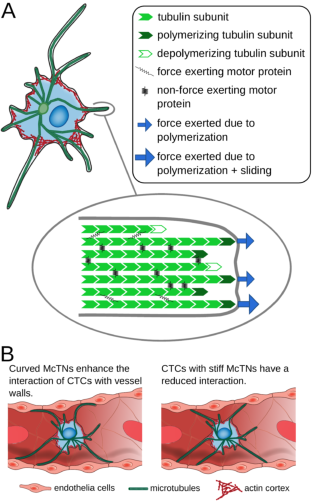}
\caption{\textbf{Scheme summarizing the mechanism of McTN formation and its function.} 
(A) Polymerization of MTs drives McTN formation. The contribution of MT sliding powered 
by molecular motors is relatively weak. (B) Schematic representation of CTC within 
a blood vessel wall. Flexible McTN can enhance the contact with blood vessel walls 
and thus facilitate adhesion.}
\label{Fig8}
\end{figure}

\subsection{Flexible, long McTNs enhance adhesion of cells to vessel walls} 
We demonstrated that McTN formation is accompanied by McTN bending due to the forces exerted by the deformed membrane on the McTN tip in the presence of the MTOC. Next, we assessed if both length and bending of McTNs are crucial factors in promoting cell adhesion to blood vessel walls. The McTN was discretized and, for generality, the adhesive interaction between the segments of the McTN and a flat surface was considered nonspecific. The McTN-wall interaction was modeled using the Lennard-Jones potential \cite{Butt03}  described as $U(r)\,{=}\,4\epsilon\,[(\sigma{/}r)^{12}{-}(\sigma{/}r)^6]$. Here $\epsilon$ is the potential well depth (representing the adhesion strength), $r$ the distance between each McTN segment and the surface of the vessel wall, and $\sigma$ the zero potential distance.  Additionally, to take into account the inherent bending of the McTN, we calculated the elastic energy $E$ of the McTN set by its structure (see {\it Materials and Methods}). The interplay of adhesion and elastic energy determines the overall curvature of the McTN. However, it should be noted that the conclusions in this section remain independent of the choice of adhesive interaction. The extent of cell-wall attachment was characterized by {\it relative wall connection},defined as the fraction of the McTN length closer than a threshold distance to the surface. Since the resulting curvature of the McTN depends on the competition of elastic and adhesion energies, the parameter$E{/}\epsilon$ was added to represent the relative stiffness. By varying the McTN length and $E{/}\epsilon$, the phase diagram of the relative wall connection was obtained in terms of the McTN length and relative stiffness. While long McTNs with small relative stiffness (i.e. either with a small elastic energy or a strong adhesion strength) show the highest relative wall connection, short McTNs with high relative stiffness have the least connection (Fig. 7A). Although McTNs of kinesore-treated cells were relatively straighter, kinks were often observed along their length (see e.g. Fig. 7B). Such kinks seem to result from defects within the MT bundle. These were included in the simulation by defining a defect position along the McTN as a finite region with significantly reduced bending rigidity. To compare the results in the presence and absence of the defect, the ratio of the wall connection was calculated ($>$1 if the relative wall connection is higher for a McTN with homogeneous bending, $<$1 otherwise). The effect of the defect site on the relative wall connection strongly depends on the position of the defect along the McTN and on its relative stiffness (Fig. 7C). For low relative stiffness, the weak point minimally affects the wall connection ratio (Fig. 7C, white colors). Only for stiff McTNs differences become visible. Thus, if the defect position is close to the cell body, the defect is rather disadvantageous for McTN-wall attachment (Fig. 7C, red colors). In this case, two relatively straight pieces are energetically more favorable than a uniformly curved McTN, although the former leads to a weaker relative wall connection. If the defect position is at an intermediate position it also results in two pieces, but this time one tends to lay on the surface and thus creates a larger wall connection compared to the McTN with homogeneous bending (Fig. 7C, light blue colors). If the defect position is close to the McTN tip there is only a small difference because the McTN is already aligned with the surface (see also Suppl. Fig. S8 for more information on the McTN shape for different kink positions and relative stiffness). When the effect of all possible defect positions was averaged it became clear that such a defect is generally disadvantageous (Fig. 7D). However, it should be noted that for an accurate quantitative prediction of the effect of kinks on the overall shape of the McTN, the imposed stresses on the McTN and the actual form of the adhesive interaction need to be considered.

\section{Discussion}
We used a combination of FRAP experiments and simulations to unravel the driving mechanism of McTN formation (Fig. 8A). Polymerization of MTs anchored in the MTOC primarily drives the generation of curved McTNs. Moreover, forces generated at the MT-bundle tip form the McTN and the reaction force exerted by the deformed membrane on the tip curves the McTN by pushing it against the MTOC. The combined effects of the inherent bending of McTNs, McTN-wall adhesion strength, and McTN length determine the extent of cell attachment to vessel walls (Fig. 8B). We note that for a more comprehensive study of the overall shape of the McTN interacting with vessel walls, the interplay of structural defects and imposed stresses on the McTN need to be considered. Recent studies have shown that the curvature and stability of MTs which experience mechanical stresses (such as when polymerizing against a membrane) are influenced by the induced lattice defects that yield in rescue sites along the MT \cite{Li23,Schaedel15}. Molecular motors moving along MTs similarly induce lattice defects \cite{Triclin21}. The interplay of such defects and polymerization against a membrane affects the overall bending of the filament. In our experiments involving kinesore treatment, the role of MT sliding was more prominent. We provided evidence that this increased relevance of sliding coincides with MT detachment from the MTOC. Such detachment enhances the possible sliding distance of MTs in anterograde and retrograde directions. To our knowledge, how kinesore affects the anchorage in the MTOC has not been previously reported. The mechanism underlying this detachment remains speculative, but possibly the increased motor activity with this treatment affects the MT lattice to such an extent that it disconnects the MTs from the MTOC. Importantly, although the predominance of sliding in McTNs of kinesore-treated cells results in longer McTNs, these structures possibly contain kinks that may lead to reduced efficiency of adhesion to blood vessel walls. Our results show that the competition of elastic energy and general (nonspecific) adhesion determines the extent of cell attachment to vessel walls. This is in line with findings that active adhesion processes might not be critically involved in the process of extravasation. Instead, it depends on active processes that facilitate transendothelial migration \cite{Strilic17}. We propose that McTNs bring CTCs into close proximity with vessel walls through nonspecific interactions with walls, which potentially triggers downstream pathways and facilitates extravasation. Further, polymerization of MTs leads to curved, long McTNs which promote attachment to blood vessel walls. Future studies are planned to experimentally validate our findings from simulation using vasculature mimicking systems and intravital microscopy.We suggest that length and bending of McTNs are key parameters to evaluate the metastatic potential of CTCs and should be considered when designing novel therapeutic strategies. 

\section{Author Contributions}

F.L., M.R.S., and L.S. designed the study. F.L. and L.K. designed the experiments; L.K. and K.K.
performed the experiments; M.R.S. and L.S. developed the model; M.R.S. and J.B. performed the
simulations; L.K. and M.R.S. analyzed the data; all of the authors contributed to the interpretation
of the results; L.K. and M.R.S. wrote the manuscript.

\section{Declaration of interests}
The authors declare no conflict of interest.

\section{Acknowledgments}
This work was funded by the Deutsche Forschungsgemeinschaft (DFG) via Collaborative 
Research Center 1027. M.R.S.\ acknowledges support by the Young Investigator Grant of Saarland 
University, Grant No.\ 7410110401. We thank Rhoda J.\ Hawkins (AIMS Ghana, University of 
Sheffield), Stefan Diez (TU Dresden), Kristian Franze (Institute for Medical Physics, 
Friedrich-Alexander-Universität Erlangen-Nürnberg, Erlangen, Germany) and Laura Aradilla 
Zapata (née Schaedel, Saarland University) for fruitful discussions.

\section{Supporting citations}
Reference \cite{Chugh17} appears in the supporting material.

\section{Disclosure}
During the preparation of this work the authors used ChatGPT 3 (OpenAI) in order to to improve readability and language. After using this tool/service, the author(s) reviewed and edited the content as needed and take full responsibility for the content of the publication.

\section{Supplementary Material}

An online supplement to this article can be found by visiting BJ Online at \url{http://www.biophysj.org}.

\section{Movie legends}

\noindent\textbf{Suppl. Movie S1: Exemplary video of EB3 comets within McTNs.} Scale bar: 10 $\mu$m.
\smallskip\smallskip\smallskip\smallskip\smallskip

\noindent\textbf{Suppl. Movie S2: Exemplary FRAP video.} The arrow indicates the 
bleached area. Scale bar: 10 $\mu$m. 
\smallskip\smallskip\smallskip\smallskip\smallskip

\noindent\textbf{Suppl. Movie S3: Microtubule dynamics upon paclitaxel treatment.} EB3 comets 
(orange) show the dynamics of growing MTs before and after treatment with paclitaxel. The addition 
of paclitaxel leads to the displacement of the cell at $t\,{\simeq}\,$123 s. Afterwards, the EB3 
comets first reduce in size and then disappear. Scale bar: 10 $\mu$m. 

\newpage

\begin{widetext}

\section{Supplemental Materials and Methods}

\subsection{Microtubule kymographs}
The kymograph was generated as previously described by drawing a segmented line with the McTN thickness over the McTN. Within the kymographs, the growth and shrinkage of MTs are observable as triangles of increased intensities. However, it should be noted that the identification of individual MT growth and shrinkage events was not possible in our analysis.
For the analysis, we again utilized Kymobutler \cite{Jakobs19}. Although the software did not detect as many events as were visually observable, the events it did detect were accurately identified and thus deemed reliable.

\subsection{Scanning electron microscopy}
For scanning electron microscopy, preparation of RPE-1 cells was modified from a protocol published in \cite{Chugh17}. Cells were treated with latrunculin A as described above to enable microtentacle formation. Afterwards, they were immobilized on a glass cover slip treated with Corning Cell Tak. The cell membranes were extracted using Triton X-100 and while simultaneously actin and MT were stabilized using phalloidin (Merck) and paclitaxel (Merck) respectively. Subsequently, 
the cells were fixed in a buffer containing glutaraldehyde and paraformaldehyde 
(Science Service). Samples were dehydrated using ethanol and hexamethyldisilazane ($98 \%$, 
Carl Roth and $\geq 99 \%$, Sigma Aldrich) as the final drying reagent. Cells were sputtered 
with $6{-}7\,$nm platinum. Images were obtained using a scanning electron microscope 
(Quanta 400; FEI, USA). Secondary electrons were detected (Everhart-Thornley detector) 
at 5 kV under high vacuum. 

\subsection{Supplemental statistical information}

\subsubsection{p-values figure 1 G and Figure S1 B}
Comparing the relative number of attached cells of untreated RPE-1 cells to RPE-1 cells treated with 1 1$\,\mu$M latrunculin A at different adhesion times or coatings. Mann-Whitney U-Test was performed.

\begin{table}[b]
\begin{tabular}{l|c} 
Adhesion Time/ Coating & p-Value \\ 
\hline
20 min & 0.002117802 \\
40 min & 0.012090025 \\
60 min & 0.323213811 \\
90 min & 0.009979962 \\
no coating & 0.016697614 \\
fibronectin & 0.324499634 \\
Pll-g-PEG & 8.57E-05 \\
\end{tabular}
\end{table}

\subsubsection{p-values figure 2 B-D}
Comapring different parameters of EB3-tip dynamics. Mann-Whitney U-Test was performed.

\begin{table}
\begin{tabular}{l|c}  
EB & p-Value \\ 
\hline
entering the McTN vs. rescue events & 0.002117802 \\
antero- vs retrograde polymerizing frequency & 0.012090025 \\
antero- vs retrograde polymerizing velocity & 0.323213811 \\
\end{tabular}
\end{table}

\subsubsection{p-values figure 4 A}
Comapring the mean intensity change between different conditons at different times. T-Test was performed.

\begin{table}
\begin{tabular}{|c|c|c|c|c|} 
time intervall & region & control vs paclitaxel & control vs kinesore & paclitaxel vs kinesore \\ 
\hline
60 s & ROI 1 & 0.060364587 & 0.246988891 & 0.278848929 \\
 & ROI* & 1.08E-07 & 3.12E-06 & 0.408139822 \\
 & ROI 2 & 0.89258815 & 0.986934693 & 0.789238834 \\
300 s & ROI 1 & 0.625437894 & 0.03451546 & 0.065166381 \\
 & ROI* & 0.00025487 & 0.000484683 & 0.822761466 \\
 & ROI 2 & 0.073386801 & 0.023769274 & 0.566889463 \\
600 s & ROI 1 & 0.466570794 & 0.000330385 & 0.012245001 \\
 & ROI* & 0.018806324 & 0.000177223 & 0.18446491 \\
 & ROI 2 & 0.019837431 & 0.001213483 & 0.191483669 \\
\end{tabular}
\end{table}

\subsubsection{p-values 6 A}
Comapring McTN straightness and McTN length for control condition, paclitaxel condition and kinesore condition. Mann-Whitney U-Test was performed because data were not normally distributed.

\begin{table}
\begin{tabular}{l|c|c|c} 
Parameter & control vs paclitaxel & control vs kinesore & paclitaxel vs kinesore \\ 
\hline
straightness & 1.74E-08 & 1.61E-23 & 1.46E-12 \\
length & 0.058759533  & 0.001120942 & 6.73E-06 \\
\end{tabular}
\end{table}

\newpage

\section{Supplementary Figures}

\begin{center}
\includegraphics[width=0.45\linewidth]{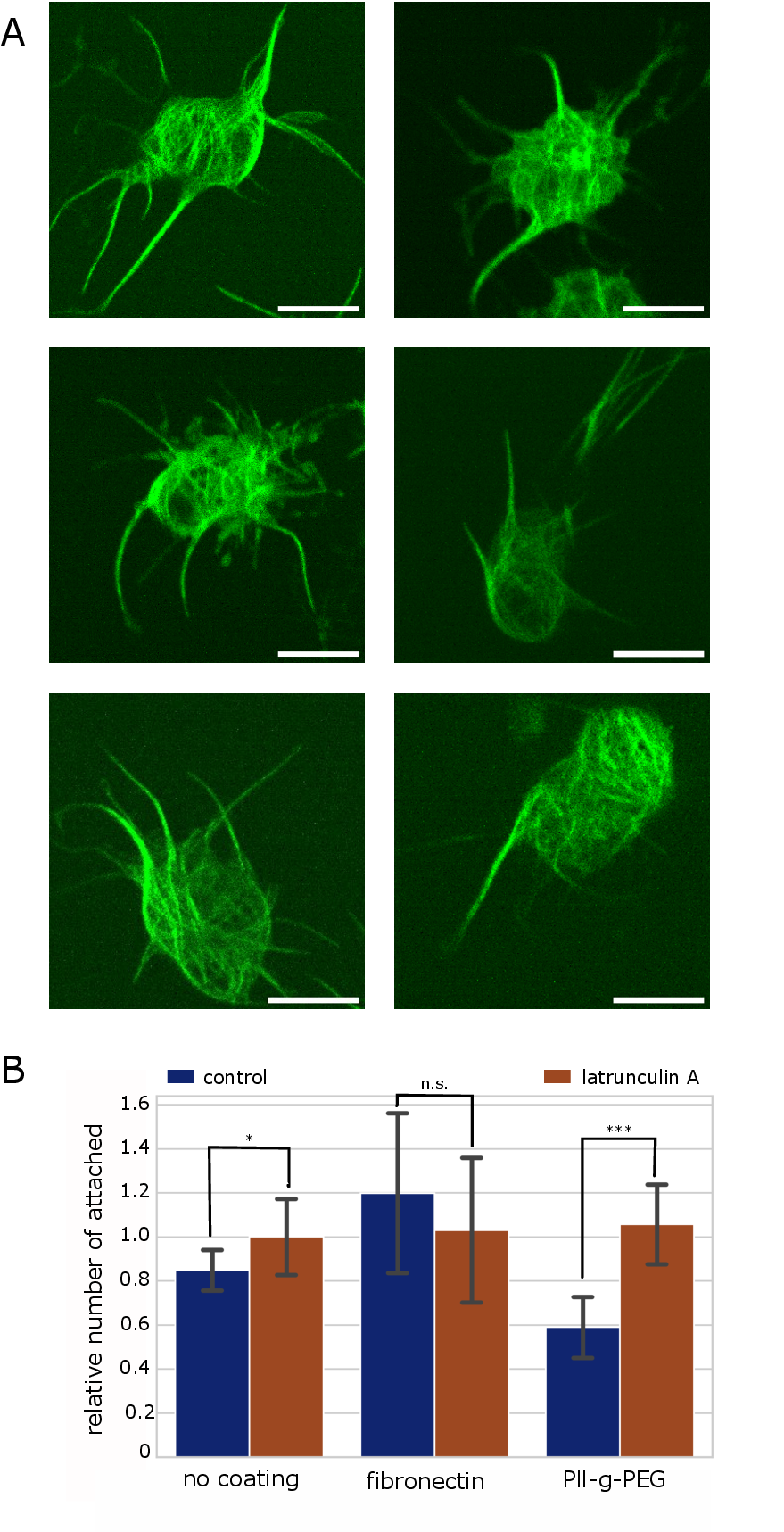}
\end{center}
\textbf{Fig. S1: McTNs in RPE-1 cells} (A) Suspended RPE-1 cells are treated with 1 $\mu$M latrunculin A form McTNs that vary in number, length and curvature. Scale bars:~10$\,\mu$m. (B) McTNs improve cell attachment independent of surface coating. RPE-1 cell attachment was tested on uncoated surface, fibronectin-coated surface and Pll-g-PEG-coated surface. Untreated cells (blue bars) are compared to cells treated with 1 $\mu$M latrunculin A (red bars) to induce McTN formation.  The relative number of attached cells was determined with a MTT assay. Attached cells after 5h 30 min serve as a reference value ($^{***}$p${\leq}0.001$,  $^{**}$p${\leq}0.01$,  $^{*}$p${\leq}0.05$, n.s., not significant; Mann-Whitney-U tests, n ${\geq}$6).
\vspace{40mm}

\begin{center}
\includegraphics[width=0.85\linewidth]{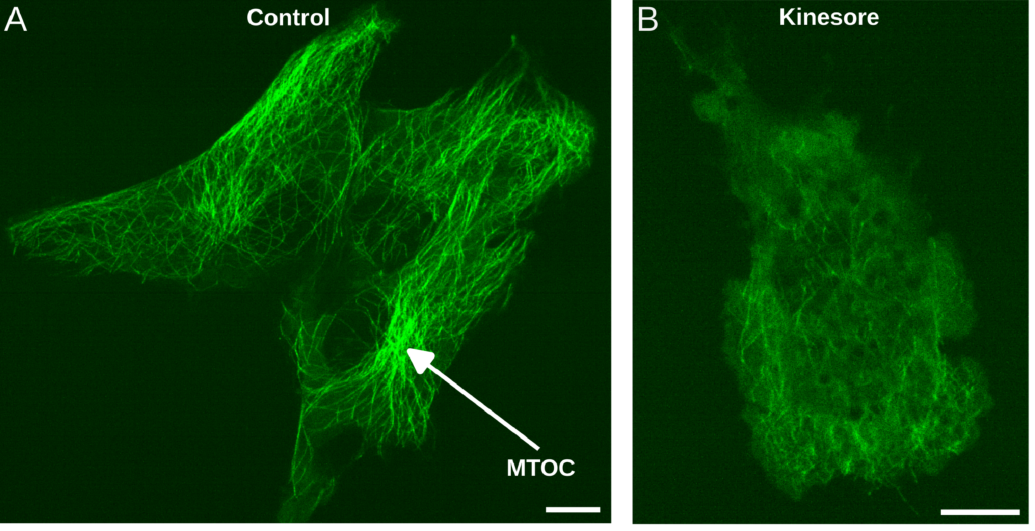}
\end{center}
\textbf{Fig. S2: Microtubule network in adhered RPE-1 cells.} (A) In untreated cells, 
the MTOC is clearly visible as a point from which MTs emerge. (B) In kinesore treated 
cells, the MTOC is not visible. Instead, MTs accumulate at the cell 
periphery. Scale bars:~10$\,\mu$m.
\vspace{40mm}

\begin{center}
\includegraphics[width=0.85\linewidth]{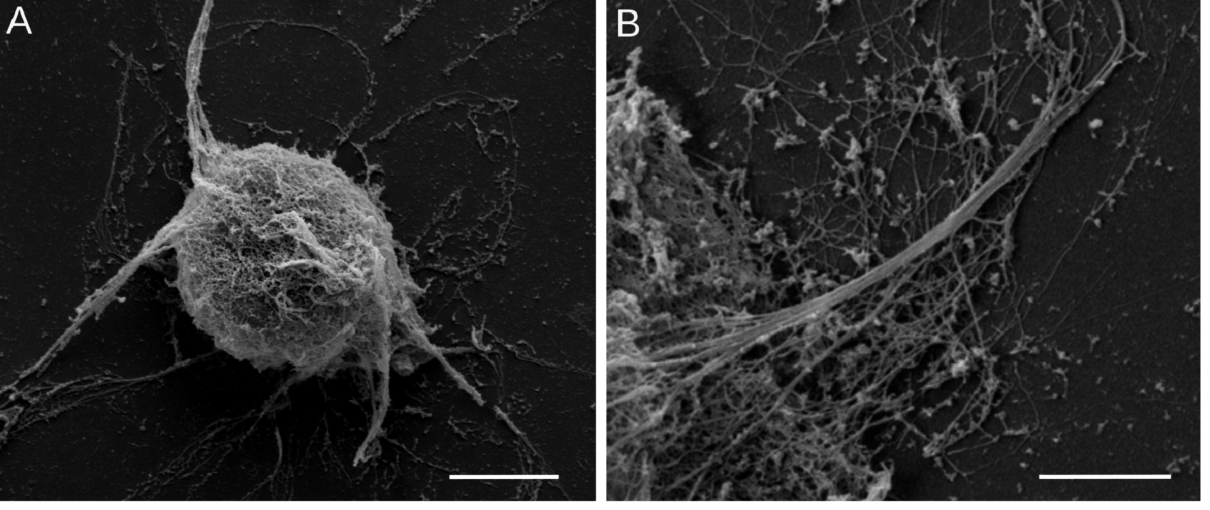}
\end{center}
\textbf{Fig. S3:\ Scanning electron images of RPE-1 cells with McTNs.} (A) Image of a 
whole cell after membrane extraction. Scale bar: 10$\,\mu$m. (B) Another cell 
is shown with a higher magnification, zoomed in on a single McTN. There is no sign of 
MT fragments inside the McTN. Scale bar: 1$\,\mu$m.

\begin{center}
\includegraphics[width=0.85\linewidth]{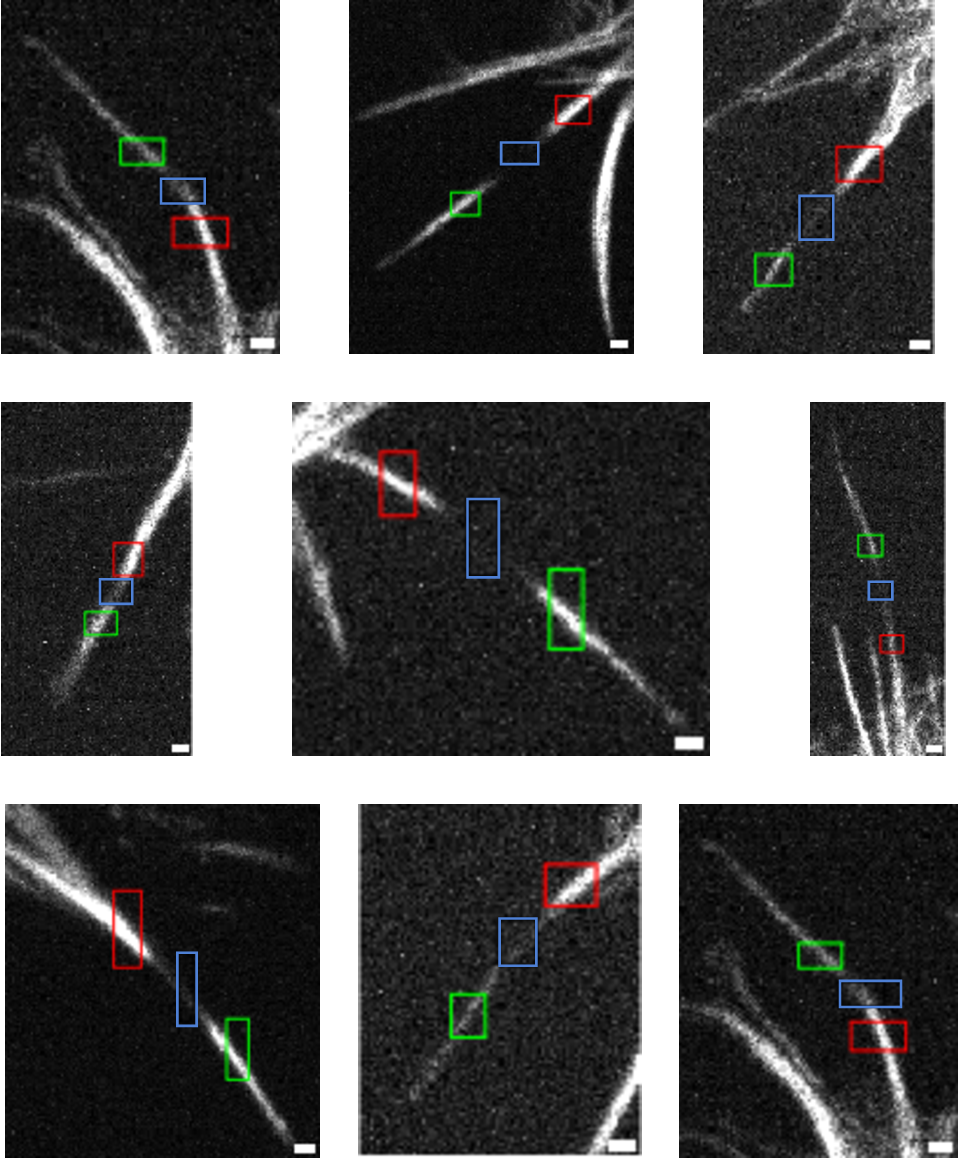}
\end{center}
\textbf{Fig. S4:\ Exemplary regions of Interest (ROIs) analyzed in FRAP experiments.} Different examples of the chosen ROIs for FRAP experiments are shown.  The bleach ROI was manually chosen to be at half length of the McTN. The size of ROIs was manually selected to cover 1$\,\mu$m of McTN. Red square: ROI1, blue square: ROI*, green square: ROI2. Scale bars: 1$\,\mu$m.

\begin{center}
\includegraphics[width=0.80\linewidth]{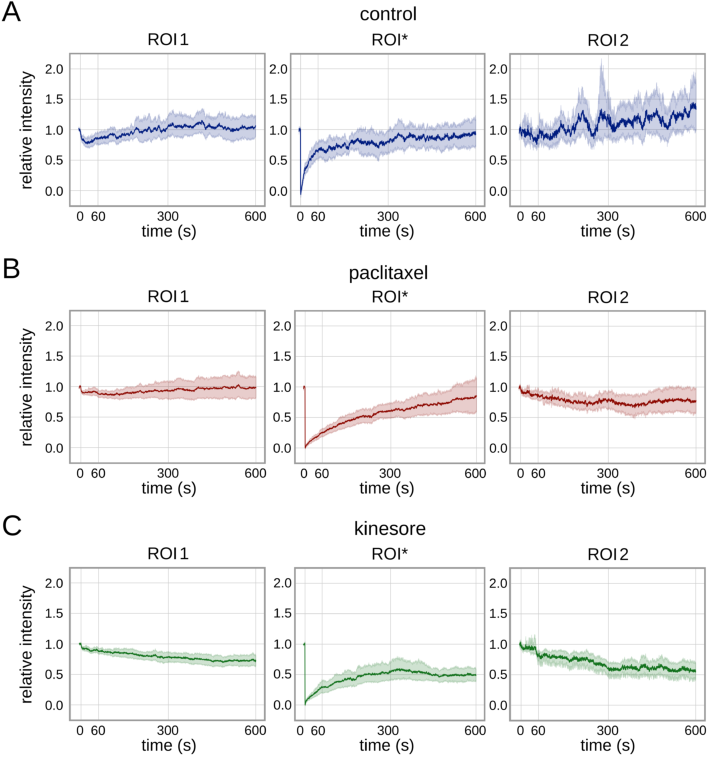}
\end{center}
\textbf{Fig. S5: FRAP results under different conditions.} Time evolution of the 
fluorescence intensity in ROI1, ROI*, and ROI2 for (A) control, (B) paclitaxel treated, 
and (C) kinesore treated cells. The mean and 95 $\%$ confidence interval is shown. (A) The intensity in the control experiments shows strong 
fluctuations which originates from the high polymerization dynamics of MTs that leads 
to different amounts of MTs inside McTNs. The intensity in ROI* first recovers very fast 
and then continues slowly. In both ROI1 and ROI2, the intensity slightly drops after 
the bleaching pulse but it falls within the fluctuation range of the data. (B) For 
paclitaxel treated cells the recovery is significantly slower. While a very slight 
drop of intensity can be seen again in ROI1, there is a clear decrease in ROI2. (C) 
For kinesore treated cells the recovery in ROI* drastically slows down. The continuous 
decrease of intensity in ROI1 and ROI2 strongly indicates sliding activity.

\begin{center}
\includegraphics[width=0.7\linewidth]{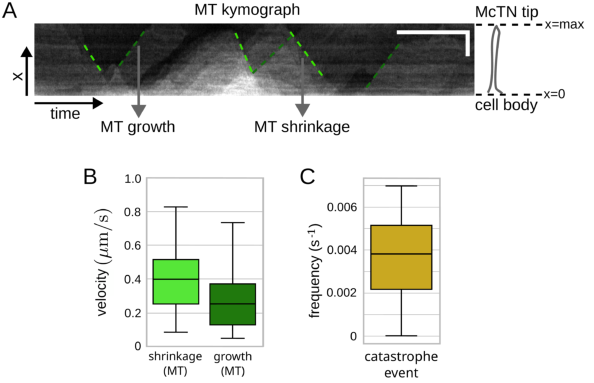}
\end{center}
\textbf{Fig. S6: Additional parameters extracted from MT and EB3 kymographs.} 
(A) Exemplary kymograph of MT-bundle signal along the McTN. Light (dark) green lines 
indicate MT depolymerization (polymerization). Scale bars:\ time, 60\,s; space, 5$\,
\mu$m. (B) Growth (polymerization) and shrinkage (depolymerization) velocities 
extracted from MT kymographs. (C) Frequency of catastrophe events extracted from 
EB3 kymographs.
\vspace{10mm}

\begin{center}
\includegraphics[width=0.85\linewidth]{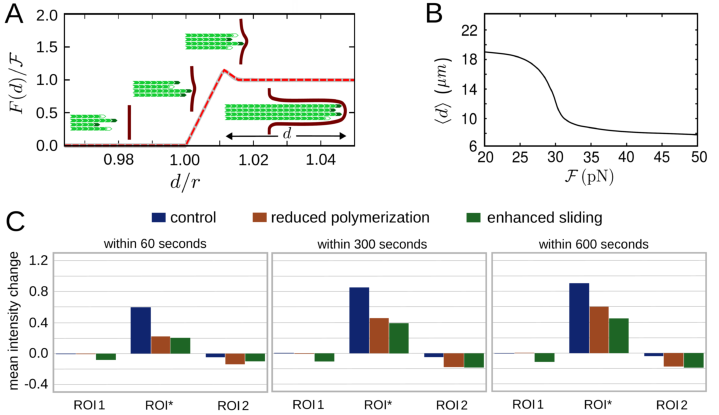}
\end{center}
\textbf{Fig. S7: MT growth and FRAP simulations}. (A) Force exerted on the MT tip 
(scaled by the asymptotic force $\mathcal{F}$) versus the MT length $d$ scaled by 
the cell radius $r$ (solid gray line). The dashed line represents the piecewise linear 
approximation used in our simulations. Insets are schematics of a MT bundle growing 
against the barrier in different deformation-force regimes. (B) Mean MT length $\langle 
d \rangle$ in terms of the asymptotic force $\mathcal{F}$ for a cell of radius 10$\,
\mu$m. (C) Mean intensity change within first 60, 300, or 600 seconds in 
ROI1, ROI*, and ROI2 obtained from control (blue), reduced polymerization (brown), 
and enhanced sliding (green) simulations. 

\begin{center}
\includegraphics[width=0.95\linewidth]{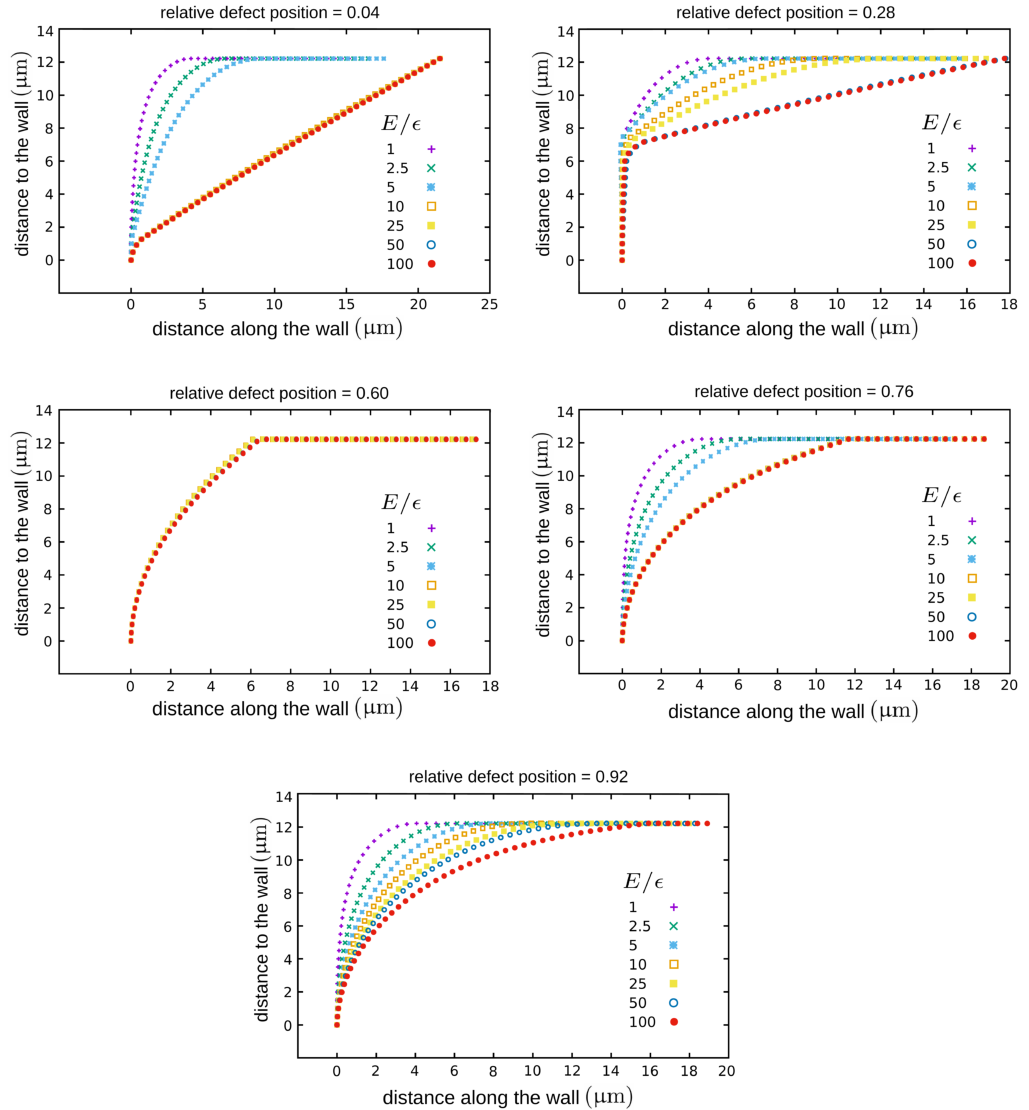}
\end{center}
\textbf{Fig. S8: Filament shape for different defect positions.} 
The positions of successive filament subunits are shown for different choices of the relative 
defect position with respect to the filament tip. In each panel, the relative stiffness of the 
filament $E{/}\epsilon$ is also varied. The filament length is 25$\,\mu$m and $E{/}
\epsilon$ is shown in units of Gm$^3$.

\end{widetext}

\bibliography{sample}

\end{document}